# Super-resolution multicolor fluorescence microscopy enabled by an apochromatic super-oscillatory lens with extended depth-of-focus

Wenli Li,[1,2,3] Pei He,[1,2,3] Yulong Fan,[4] Yangtao Du,[5] Bo Gao,[6] Zhiqin Chu,[7] Chengxu An,[1,2,3] Dangyuan Lei,[4],* Weizheng Yuan[1,2,3],* and Yiting Yu[1,2,3],*

[1]Research & Development Institute of Northwestern Polytechnical University in Shenzhen, College of Mechanical Engineering, Ningbo Institute of Northwestern Polytechnical University, Northwestern Polytechnical University, Xi'an 710072, China

[2]Key Laboratory of Micro/Nano Systems for Aerospace (Ministry of Education), Northwestern Polytechnical University, Xi'an 710072, China

[3]Shaanxi Province Key Laboratory of Micro and Nano Electro-Mechanical Systems, Northwestern Polytechnical University, Xi'an 710072, China

[4]Department of Materials Science and Engineering, City University of Hong Kong, Hong Kong 999077, China

[5]The Institute of AI and Robotics, Fudan University, Shanghai 200433, China

[6]Key Laboratory of Spectral Imaging Technology of Chinese Academy of Sciences, Xi'an Institute of Optics and Precision Mechanics, Chinese Academy of Sciences, Xian 710119, China

[7]Department of Electrical and Electronic Engineering, Joint Appointment with School of Biomedical Sciences, The University of Hong Kong, Hong Kong 999077, China

*Email: dangylei@cityu.edu.hk; yuanwz@nwpu.edu.cn; yyt@nwpu.edu.cn

**Abstract**: Multicolor super-resolution imaging remains an intractable challenge for both far-field and near-field based super-resolution techniques. Planar super-oscillatory lens (SOL), a far-field subwavelength-focusing diffractive lens device, holds great potential for achieving sub-diffraction-limit imaging at multiple wavelengths. However, conventional SOL devices suffer from a numerical aperture (NA) related intrinsic tradeoff among the depth of focus (DoF), chromatic dispersion and focus spot size, being an essential characteristics of common diffractive optical elements. Typically, the limited DoF and significant chromatism associated with high NA can lead to unfavorable degradation of image quality although increasing NA imporves the resolution. Here, we apply a multi-objective genetic algorithm (GA) optimization approach to design an apochromatic binary-phase SOL that generates axially jointed multifoci concurrently having prolonged DoF, customized working distance (WD) and suppressed side-lobes yet minimized main-lobe size, optimizing the aforementioned NA-dependent tradeoff. Experimental implementation of this GA-optimized SOL demonstrates simultaneous focusing of blue, green and red light beams into an optical needle of $\sim0.5\lambda$ in diameter and $>10\lambda$ in length (DoF) at 428 $\mu$m WD, resulting in an ultimate resolution better than $\lambda/3$ in the lateral dimension. By integrating this apochromatic SOL device with a commercial fluorescence microscope, we employ the optical needle to perform, for the first time, three-dimensional super-resolution multicolor fluorescence imaging of the "unseen" fine structure of neurons at one go. The present study provides not only a practical route to far-field multicolor super-resolution imaging but also a viable approach for constructing imaging systems avoiding complex sample positioning and unfavorable photobleaching.

Optical microscope, a basic tool for exploring and revealing the "secrets" of life science at the cellular and subcellular levels, often suffers from the resolving limit defined by the Abbe-Rayleigh diffraction limit. Ever-growing efforts have been dedicated to overcoming the diffraction limit, and relevant methods can be categorized into the near-field label-free and the far-field fluorescent-labeling approaches based on their respective working principles. In the near-field regime, the negative-index superlens,[1, 2] field concentrators,[3-5] microsphere-assisted imaging[6, 7] and scanning near-field optical microscopy (SNOM)[8] can resolve the fine details of an object located tens of nanometers from the lenses by invoking and collecting high spatial frequencies of evanescent waves. However, it has been difficult to implement these approaches for practical applications in many scenarios, such as biological imaging, mainly due to the extremely small working distances (WDs). In the far-field regime, stimulated emission depletion (STED) microscopy,[9, 10] photo-activated localization microscopy (PALM),[11, 12] and stochastic optical reconstruction microscopy (STORM)[13, 14] have been demonstrated to achieve super-resolution imaging at WDs of tens of micrometers. But these far-field super-resolution imaging methods are only applicable for labeled biological specimens, and often require the use of high-energy lasers, mercury or xenon lamps to selectively activate labeled fluorophores or single molecules, resulting in considerable photobleaching, fluorescence quenching and irreversible photodamage. In addition, some of these approaches involve complicated post-processing algorithms to obtain computed images that overcome the fundamental resolution limit of the optical lenses used. Notably, multicolor super-resolution imaging can enable spectral visualization of molecular interactions in biological samples,[15-17] but it still remains an intractable challenge for the above-mentioned far-field and near-field imaging techniques due to issues like chromatic aberration and limited depth-of-focus (DoF).

Recently, a non-invasive, label-free, far-field super-resolution imaging technique, called super-oscillatory lens (SOL) optical microscopy, has been proposed to address these issues.[18-

[23] Nevertheless, the SOL devices developed thus far still suffer from the following limitations: 1) short DoF resulting from highly compressed light fields by high numerical aperture (NA);[18, 24-26] 2) small field of view (FoV) limited by the strong side-lobes;[27, 28] 3) chromatic aberration arising from the wavelength-dependent phase delay;[18, 29-31] and 4) short WDs at the level of tens of micrometers.[24, 26, 29-32] Many efforts have been devoted to surmounting these inherent limitations and the corresponding results are summarized in Tab. S1. Compared with conventional single-hotspot focusing patterns,[18, 24-31] the central stop of phase plate[19, 20] and the vectorial polarization beams[21, 22] are applied to extend the SOL DoF. Unfortunately, the former method sacrifices the focusing efficiency and degrades the imaging resolution in the transverse plane, and the latter can only yield hollow needles with DoF≤10λ. Consequently, the best apochromatic two-dimensional (2D) planar SOL reported in the literature has a 1~2 µm long DoF and an extremely low focusing efficiency of less than 3%, making it unsuitable for realistic bio-imaging applications.[18] Although three-dimensional (3D) super-resolution imaging has recently been demonstrated with a DoF-extended SOL composed of non-subwavelength-sized chromium belts,[32] its WD, transmission efficiency and monochromaticity all need significant improvements for realistic cellular imaging applications.

To the best of our knowledge, the recently developed SOLs[26-37] can only address one or, at most, two of the four issues mentioned above. Grounded on our earlier studies,[36, 37] here, we apply multi-objective genetic algorithm (GA) to address all the four issues simultaneously by devising an apochromatic binary-phase SOL with extended DoF, increased WD and multicolor super-resolution capability. This approach achieves successive focusing in the longitudinal direction, forming an optical needle with a DoF larger than 10λ, which is orders of magnitude longer than the best results reported in the literature.[18, 36] Multicolor super-resolution imaging can hence be realized by overlapping the optical needles of blue, green and red light beams, with a focusing efficiency of 11.2% at a focal distance of 428 µm, far surpassing the results in Ref. 24 and other reports.[26-37] Finally, we showcase the application of the developed SOL on a fluorescence microscope for label-free imaging, which reveals a far-field resolution limit of 0.3λ at 488 nm under continuous scanning. We demonstrate, for the first time, the multicolor fluorescence 3D imaging of labeled neuron tissues at one go. Such a versatile, lightweight and cost-effective high-NA SOL-based microscope will find promising applications in the multicolor super-resolution 3D imaging of label-free inorganic materials and labeled organic tissues, and thus significantly extend the scope of optical microscopy to the regimes that cannot otherwise be achieved by the commercially available confocal, multiphoton, STED, PALM and STORM microscopies. Importantly, the apochromatic SOL with extended DoF can also be applied in diverse fields, including super-resolution multicolor stereomicroscopy,[38] tunable spectral microscopy,[39] and targeted therapy beyond the diffraction limit.[40]

**RESULTS**

The DoF of a conventional diffractive lens follows DoF=$\lambda/[n(1-\cos\alpha_{max})]$, dictated by the optical analog of the uncertainty principle.[41, 42] The largest convergence angle $\alpha_{max}$ between the outermost ray and the optical axis relates to the maximum spatial frequency of an imaging system with NA=$n \sin\alpha_{max}$. As a result, a larger NA gives rise to a compressed focus spot (i.e. a smaller spot size) accompanied with a reduced DoF, as depicted in the middle and right-hand panels of Fig. 1a. On the other hand, the increase of the NA of a diffractive lens also incurs more significant chromatic abberation due to the increased dispersion associated with the reduced operational bandwidth (see the left-hand panel in Fig. 1).[43] Therefore, there always exists an NA-dependent intrinsic tradeoff between the DoF, focus spot size and chromatic abberation in conventional diffractive optical elements as well as SOLs. To optimize this tradeoff for maximizing the NA-dependent DoF and simultaneously minimizing the chromatic dispersion and focal spot size, here we apply a new multi-objective optimization strategy to design an apochromatic binary-phase SOL having an extended and highly uniform optical-needle-like focusing region with suppressed side-lobes. To realize such a multicolor SOL device, we first extend the DoF based on an axially joint multifoci approach and minimize the side-lobe intensity as well as the main-lobe FWHM of the focusing needle for individual wavelengths, as examplified by the blue light shown in Fig. 1b. Then, the sub-diffraction needle-like optical patterns generated from three separate wavelengths (e.g. blue, green and red) are spatially overlapped with each other to give rise to an apochromatic SOL with an extended DoF and minimized FWHM, as depicted in Fig. 1c.

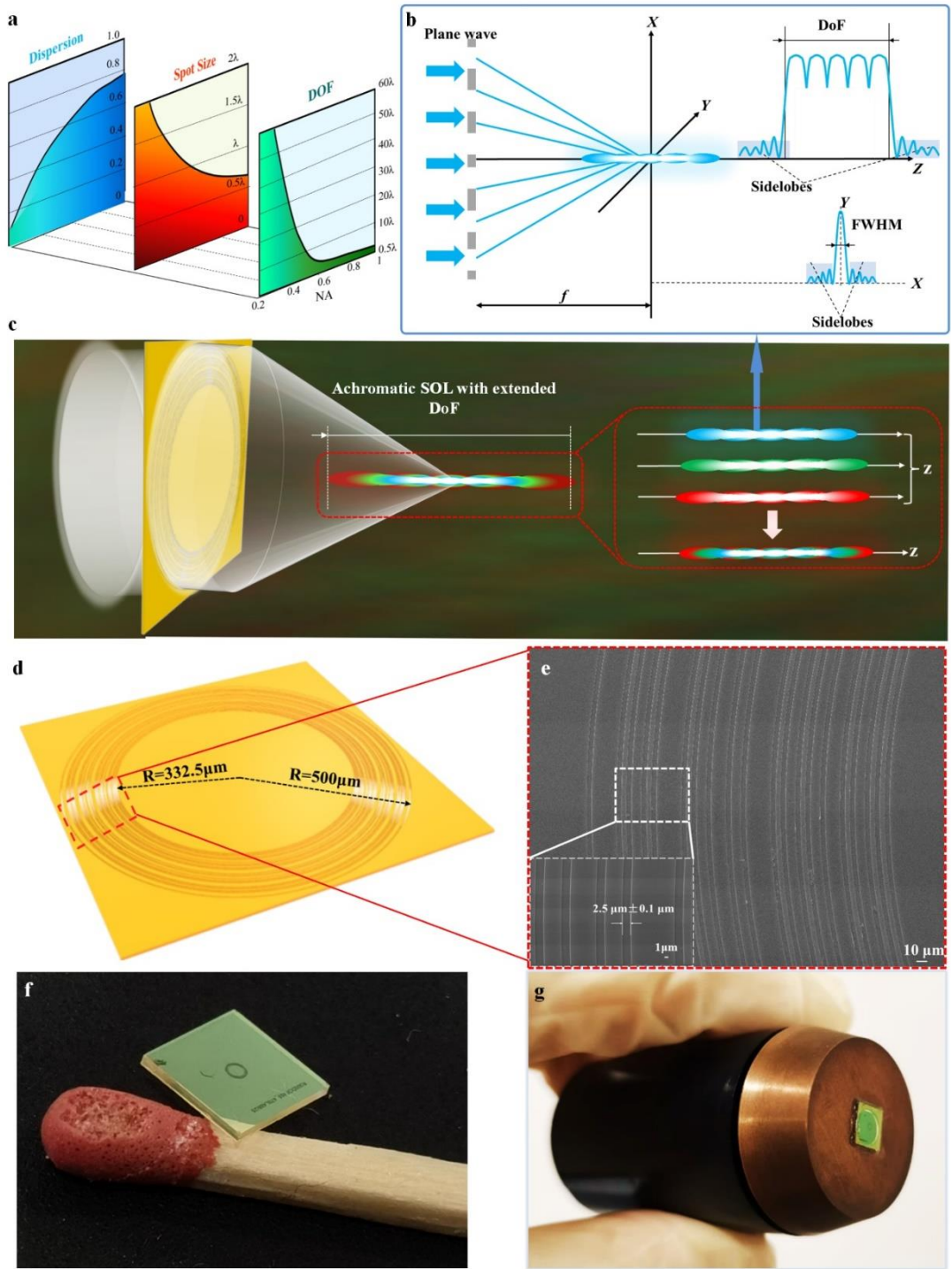

**Fig. 1**. **Design and realization of the proposed SOL**. a) NA-dependent tradeoff among the chromatic dispersion, focal spot size and DoF of a diffractive lens such as a planar SOL. b) Conceptual formation of a needle-like optical pattern by connecting adjacent multifoci under monochromatic illumination, with minimized side-lobe intensities and main-lobe FWHM. c) Schematic illustration of sub-diffraction-limit focusing by an apochromatic SOL with extended DoF over the whole visible spectrum. The inset sketches the formation of an apochromatic sub-diffraction-limit optical needle by superposing the focus contours at the blue, green and red light wavelengths. d) Schematic illustration of the proposed apochromatic SOL device drilled in a 215-nm-thick $Si_xN_y$ film. The innermost and outermost rings are 332.5 μm and 500 μm in radius, respectively. e) Scanning electron microscopy (SEM) micrograph of the device with a zoom-in view shown in the inset.

f) Optical image of the fabricated SOL compared with a matchstick. g) Optical image of a customized SOL-based objective.

As presented by Eqs. M1 and M2, we consider a universal optimization model to construct an achromatic optical pattern with an extended DoF (See **Methods, GA-based optical field design** for design details). The GA process is triggered by initializing hundreds of individuals with random design parameters as the initial generation. Then, the individuals are screened by the objective functions G1, G2, and G3 in Eq. M1, serving as the fitness functions for optimizing the DoF, the focusing spot intensity uniformity and the main-lobe FWHM at each working wavelength, respectively, which are then weighted and summed up to formulate the multi-objective optimization to purposely address the abovementioned intrinsic tradeoff faced by conventional SOLs. After that, screened individuals mutate and cross-over to generate the offspring. The GA loop will continue if the ending condition (i.e. the maximum iteration step set in the optimization process) is not fulfilled; Otherwise, the loop will end up with an optimal optical pattern and return to us with the corresponding phase mask. Detailed optimization iteration processes and corresponding results can be found in Supplementary Information Note 1 and Figs. S1-S3.

The phase mask obtained by the GA optimization process is given in Fig. S2. The designed SOL composed of concentric rings with varying radii perforated in a 215-nm-thick $Si_xN_y$ film conforming to the result of Fig. S2, is verified by the SEM image, as shown in Fig. 1e. The millimeter-scale device was fabricated through the conventional optical lithography process (see Methods, Wafer level fabrication and Fig. S4). Notably, the fabrication accuracy can be controlled at 0.1 μm by the standard UV lithography steps, ensuring the required focusing performance of SOLs with minimal deviation from the set design. The fabricated SOL compared with a common matchstick is displayed in Fig. 1f to demonstrate its millimeter-scale aperture. And the configuration of the fabricated SOL integrated onto a customized objective is revealed in Fig. 1g.

The theoretical calculation of a conventional diffractive lens with the same NA (=0.76) gives the results for the DoF of 1.4 μm at $\lambda_B$ = 488 nm, 1.5 μm at $\lambda_G$ = 532 nm, and 1.8 μm at $\lambda_R$ = 640 nm, all of which are inferior to 3λ. To characterize the far-field focusing properties, fiberized-coupled lasers at wavelengths of 488 nm, 532 nm and 640 nm with a linearly polarized beam are used to illuminate the customized SOL from the substrate side. A Nikon inverted microscope was used to capture the light field (See Supplementary Information Note 3 and Fig. S5 for the details of experimental validation of far-field focusing properties). The measured elongated hotspot within the range 422.9~430.1 μm at $\lambda_B$ = 488 nm, overlaps with the hotspot range 425.1~432 μm at $\lambda_G$ = 532 nm and 425.6~432.2 μm at $\lambda_R$ = 640 nm, clearly revealing the significant DoF extension of our elaborately designed SOL. The measured focal lengths of $z_f \approx$ 428 μm in Fig. 2b,d,f at the three wavelengths (488 nm, 532 nm and 640 nm) correlate well with the simulation results presented in Fig. 2a,c,e, validating the chromatic-dispersion-free property of our designed SOL. Furthermore, the corresponding three-dimensional (3D) views of the simulated and measured transverse intensity profile at $z_f$ = 428 μm for the design wavelengths are respectively illustrated in Fig. 2g,j,m,h,k,n. The measured transverse modes verify the simulated results, implying that the intensity pattern is dominated by the central main hotspot without significant side-lobes. To quantitatively characterize the spot size of the created optical patterns, we plot both the simulated and measured intensity profiles at $\lambda_B$, $\lambda_G$ and $\lambda_R$ in Fig. 2i,l,o, where good agreement is revealed between the simulated and measured results at $\lambda_G$ = 532 nm, and a slight difference appears for the remaining two wavelengths. The dissimilarity between the measured and simulated results is probably caused either by the imperfect integration of the proposed SOL with the optical objective, or due to the lack of precision collimation between the objective and the incident optical excitation. More precise micro-electromechanical systems (MEMS) technics may further improve the focusing quality at the aforementioned two wavelengths.

We carry out a parametric study of the theoretical, simulated and measured results in Tab. S2 to provide a direct proof of the advanced characteristics of our optimized SOL. All the experimental results are the average of three measurements. Here, we define a figure of merit $V = DoF_{Sim}/DoF_{Theory}$ to quantify the DoF extension. An extension of at least 3.7 times for all three wavelengths is found, thus demonstrating the prominent DoF extension effect and the robustness of our design approach. We also list the simulated and measured full width at half maximum (FWHM) of the hotspots at three wavelengths: 245 nm ($0.502×\lambda_B$), 270 nm ($0.507×\lambda_G$) and 360 nm ($0.562×\lambda_R$) based on the simulated results and 256 nm ($0.525×\lambda_B$), 270 nm ($0.507×\lambda_G$) and 390 nm ($0.609×\lambda_R$) based on measured results, which significantly break the Abbe diffraction limit ($0.66\lambda$) with NA=0.76. Other characteristics of our customized SOL with the focusing efficiency over 11.2% at three wavelengths are additionally presented, both of which surpass the best reported apochromatic SOL,[18] as given in Tab. S2.

Moreover, we plot the curves of the measured FWHMs at different axial distances to investigate the uniformity of the designed optical needle-like focusing pattern. As demonstrated in Fig. 2p, the FWHMs break through the diffraction limit at five typical points within the focusing region at discrete axial distances which are extracted and marked as red asterisks in bold for the three incident design wavelengths (see Supplementary Information Note 4 and Fig. S6 for more details).

The optical super-oscillation is the result of the delicate interference of far-field propagating waves, which can be accurately engineered to achieve sub-diffraction-limit foci preserving the fine physical details of objects.[44, 45] The propagation region of interest (RoI) is colored within the range of $0.38\lambda/NA$ and $0.5\lambda/NA$ to highlight the DoF extension of super-resolution focusing at different incident wavelengths. Although there are some intensity fluctuations within the optical needle, the optical energy is confined within the sub-diffraction-limit region for all the three incident wavelengths. Taking $\lambda_G$ as an example, five typical points at discrete propagation distances from z = 425 μm to z = 432 μm at a step of 1 μm are extracted, and the corresponding FWHMs measured are evaluated as 0.4511$\lambda_G$, 0.5075$\lambda_G$, 0.5075$\lambda_G$, 0.5075$\lambda_G$ and 0.5075$\lambda_G$, respectively, all of which break the Abbe diffraction limit. Henceforth, the focal spot sizes and contours are quasi-uniform within the customized needle-like focal region from z = 425 μm to z = 432 μm. This could greatly benefit the practical bio-imaging applications.

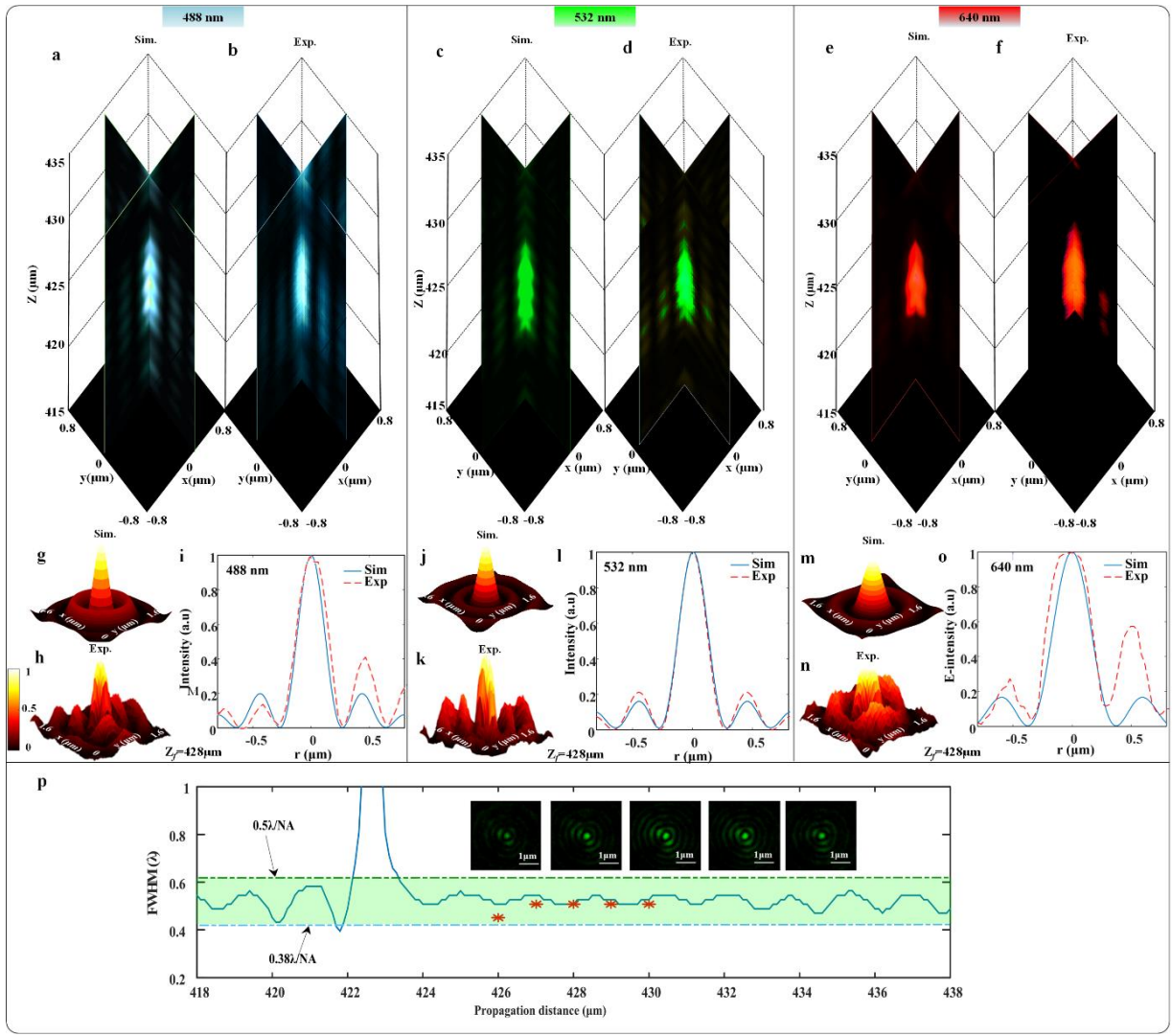

**Fig. 2**. **Apochromatic far-field subwavelength focusing.** a,c,e) Simulated and b,d,f) experimental field patterns (in combined x-z and y-z longitudinal sections) at $\lambda_B$ = 488 nm a,b), $\lambda_G$ = 532 nm c,d), and $\lambda_R$ = 640 nm e,f) by an apochromatic SOL device with extended DoF centered at $z_f$ = 428 μm. g,j,m) Simulated and h,k,n) measured intensity profiles in the transverse focal planes for the patterns in a-f). Exp., experiment; Sim., simulation. i,l,o)

Comparison of the simulated and measured intensity profiles in the radial direction at $z_f = 428$ μm. p) Theoretical (solid) and measured (red asterisk) FWHM values in the propagation direction for $\lambda_G = 532$ nm. The insets show the measured transverse intensity distribution profiles in an area of $3.5 \times 3.5$ μm$^2$ from z = 426 μm to z = 430 μm with a step size of 1 μm.

**2D sub-diffraction multicolor imaging.** Based on the optimized apochromatic SOL with extended DoF, we establish a customized high-resolution multicolor microscope system for the scanning mode of high-quality imaging applications with a large FoV. To test its resolving capability, a homemade nanometer-scale resolution target is utilized. The imaging target is fabricated on a 100-nm-thick chromium layer by focused ion beam (FIB) milling. The working principle of the customized imaging system is sketched in Fig. 3a, in which the apochromatic SOL is used to provide the sub-diffraction-limit needle-like optical contour along z axis. Piezoelectric transducer provides precise mechanical movement to ensure the relative displacement between the sample and the light needle (see Supplementary Information Note 5 and Fig. S7 for the imaging setup and the process). The center to center (CTC) distances of the resolution chart are set as 150 nm, 250 nm and 350 nm, respectively, as shown in Fig. 3b. Noting that the CTC distance of the smallest etched gap is beyond the diffraction limit for the visible light. This pattern is examined under a conventional inverted microscope (objective NA=0.9) without the SOL in the transmission mode. As a result, the collected signals of the isolated gaps are mapped in the vague appearances for the three incident wavelengths where the slits cannot be optically resolved (Fig. 3c-e). In comparison, our apochromatic SOL-based microscope is able to reveal the fine topology of the nanoscale transparent slits, even the 150-nm-width gap when illuminated at the wavelength of 488 nm (Fig. 3f). To alleviate the effect of sidelobe intensity on the imaging results and simplify the whole setup, the sidelobe energy ratio at the focal plane is purposely set below 0.3 compared with the central focal spot in the GA process. During the imaging process, the whole images within the FoV for the objective are collected by the built-in charge-coupled device (CCD) camera. To reconstruct the final complete images, the template matching algorithm is applied and then the serial images are merged together. The ultimate reconstructed images of the slits from 150 nm to 350 nm at the three incident wavelengths (Fig. 3f-h) also show that our SOL-based microscope could achieve a high-quality image and reveal the objects with smaller distortion at a low cost. In Fig. 3i-k, a quantitative comparison between the line-scanning intensity profiles confirms that our SOL-based microscope exhibits a CTC resolution of 150 nm (i.e., 0.30$\lambda$ at $\lambda$=488 nm). Three apparent contrasts between the peak and valley of the extended-DoF SOL profile ranging between 40% and 50% for the three wavelengths can be observed. This value is relatively higher than the 20% as defined by Rayleigh resolution criterion for distinguishing two neighboring points in one image.[46] Based on the point scanning imaging mode, narrower slits etched in the metallic layer will be resolved via the customized SOL-based optical microscopy.

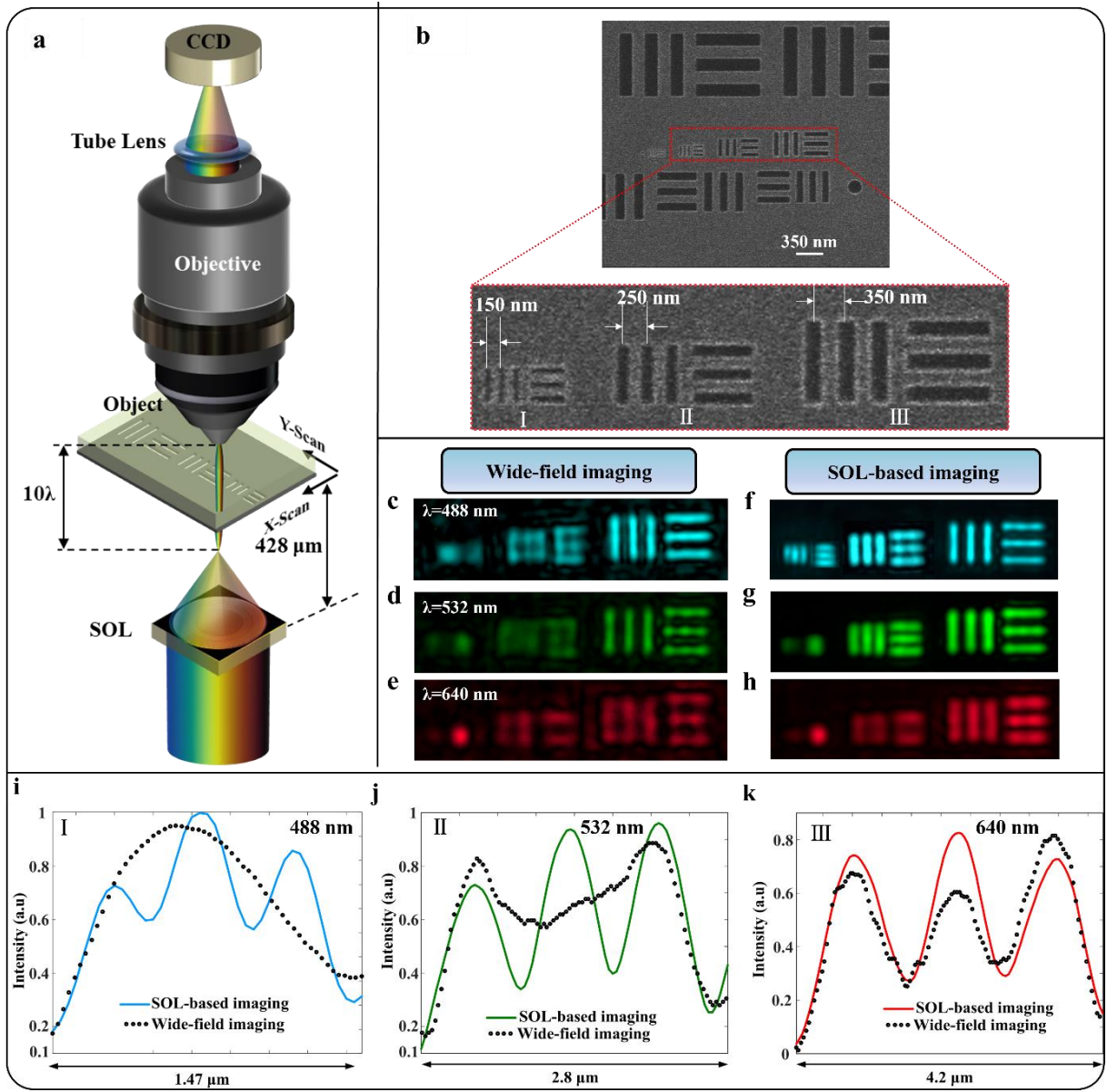

**Fig. 3**. a) Sketch of an SOL-based scanning optical microscope. When the SOL is illuminated by an incident laser, a sub-wavelength-focused 10λ-long optical needle is formed at 428 μm away from the SOL output plane. The imaging sample is placed within the needle length range and scanned in the x-y plane. The transmitted light is collected by an objective and sent to a CCD camera for imaging. b) SEM micrographs of the resolution testing chart. The center-to-center (CTC) distances between the slits in structures (I-III) are 150 nm, 250 nm, and 350 nm, respectively. c-e) Experimental wide-field imaging and f-h) SOL-based super-resolution imaging results for the testing chart in b). i-k) Line-scanning intensity profiles of the experimental images by addressing the CTC distances from 150 to 350 nm at an interval of 100 nm. The resolvable CTC distances are 320 nm for the traditional wide-field microscope and 150 nm for our SOL-based microscope at the incident wavelength $\lambda_B = 488$ nm.

**3D dual-color cellular imaging.** Imaging biological samples with a certain thickness is usually limited by an inevitable obliquity between the sample and lens plane. This leads to an out-of-focus blurred images and therefore limits the FoV of imaging systems with a short DoF. When observing the internal fine structures of neurons where internal synapses always lie at different depths, several key points such as a high spatial resolution, an extended DoF and apochromatic property are required. Mapping the activity of neurons at high resolution can provide a great insight into how neuronal ensembles collectively drive brain function for researchers. To promote the low-cost multicolor super-resolution stereo fluorescent imaging application of the SOLs, the apochromatic DoF-extended SOL-based microscope is employed to observe the human neurons cultured and fixed on the 170-μm-thick glass slide. The cell culture and fluorescent labeling information are available in the Supplementary Materials section. The excitation and emission wavelengths of the two fluorescent dyes are 488 nm, 520 nm and 546 nm, 580 nm respectively (see Supplementary Information Note 6 and Fig. S8 for the details of 3D dual-color cellular imaging). The movie cucoloris of the imaging process under the 488 nm illumination can be found in Supplementary Materials. Fig. 4a displays the cellular image captured by the bright-field microscope, and the RoI region is highlighted in the red dotted box. The wide-field fluorescent images are showcased in Fig. 4b,c and the zoom-in figures of the wide-field fluorescent images are demonstrated in Fig. 4e,f. Note that the short DoF of the commercial imaging objective is not able to cover the neuronal synapse at different depths hence the whole fine details within RoI can only be obtained through several out-of-plane movements of the objectives. This multiple positioning will incur inevitable photobleaching and make the imaging process more complicated. Compared with the wide-field fluorescent results, the blurred morphologies can be distinctly mapped by the SOL-based microscope for the two incident wavelengths, as shown in Fig. 4h,i. To characterize the multicolor fusion results, the two images obtained at two incident wavelengths are merged, as shown in Fig. 4d,g. The 3D information of the neurons at different depths can be observed more clearly at one go by in-plane scanning of the SOL. The detailed splicing process is introduced in **Methods, Splicing of fluorescent-labeled cell images**. The final cellular imaging results demonstrate that our customized apochromatic SOL-based optical microscopy (SOM) can realize the far-field multicolor sub-diffraction-limit cellular imaging with a large DoF.

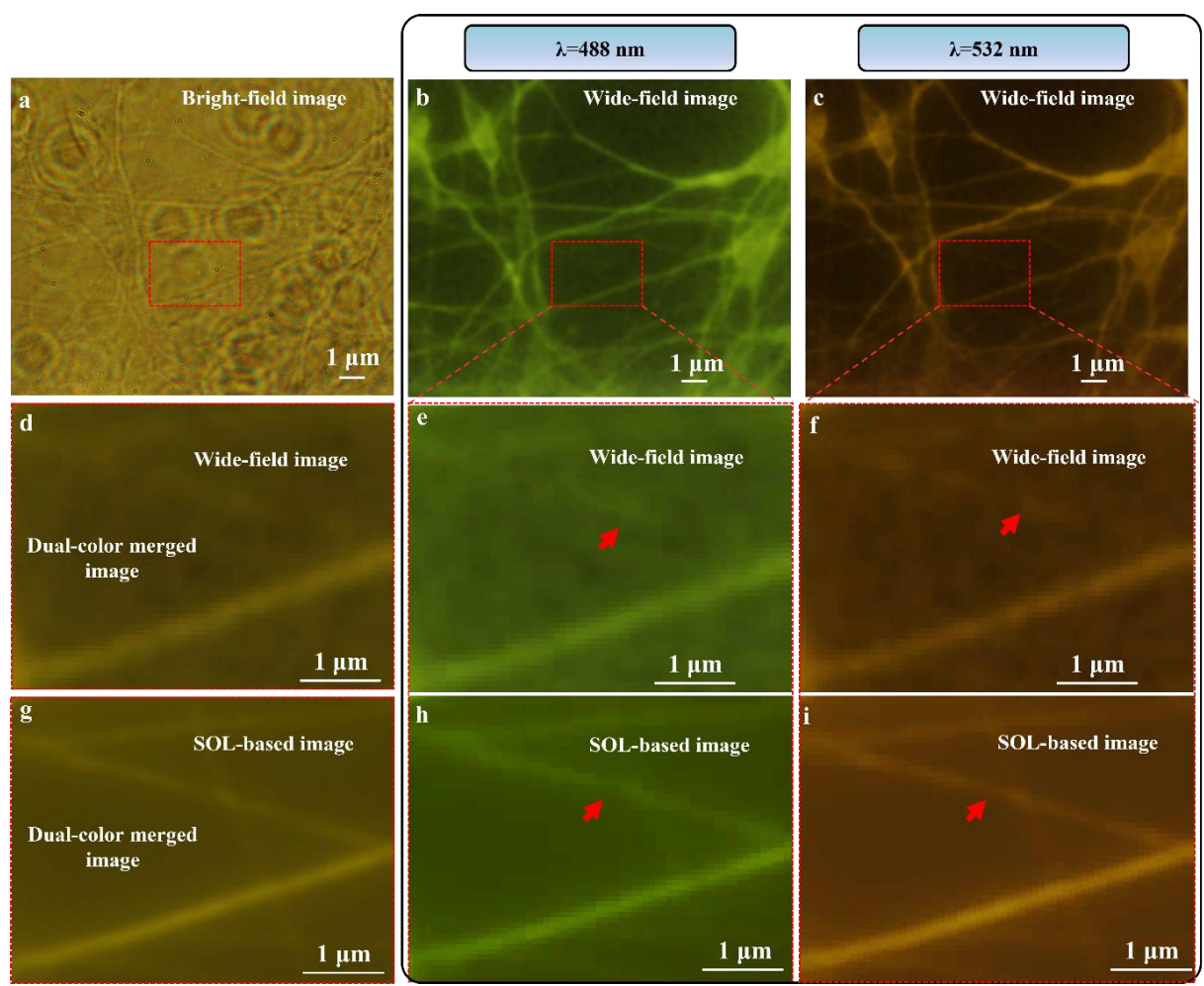

**Fig. 4**. **3D dual-color cellular imaging.** a) Bright-field and b,c) Wide-field fluorescence microscopy images of a thick neuron tissue. e,f) Zoom-in views of the enclosed areas in b,c). h,i) SOL-based microscopy images of the enclosed area in b,c). d,g) Merged dual-color fluorescence images from wide-field imaging and SOL-based microscope.

**DISCUSSION**

For the real-life high-resolution multicolor fluorescent imaging applications, the sub-diffraction-limit dispersion-corrected hotspot with an extended DoF is needed to realize the clear and informative imaging results. SOL-based optical microscope is a highly customized, non-invasive and universal imaging technique without complex mathematical post-processing. To develop the high-NA apochromatic SOL with extended DOF from the perspective of structural innovation and algorithm improvement, the axially jointed multifoci method is proposed to achieve this goal. Based on the proposed method, the SOL (NA=0.76) with extended DoFs and a same WD for three discrete incident wavelengths is optimized. The later experimental measurements demonstrate a good correlation with the simulated results for the proposed SOL. As expected, both the designed and measured DoFs exceed the theoretically results compared with the conventional diffractive lenses with the same NA. Furthermore, the ultra-long and identical WD surpasses the best reported result in literature. Finally, all the FWHMs of hotspots at three wavelengths successfully break the Abbe diffraction limit. As a result, all these outstanding features of proposed SOL allow its integration into a customized microscope to perform super-resolution multicolor imaging applications.

Then we exploit our SOL-based multicolor fluorescent microscope in the practical label-free and labeled imaging capability. Firstly, the resolving power is tested by the customized nanoscale resolution target. At the incident wavelength of 488 nm, the 150-nm-width nanoslits can be easily distinguished, in contrast to the blurred images formed by the wide-field imaging microscope. To demonstrate the practical imaging ability with the extended DoF, the apochromatic SOL is employed to observe human neurons labeled by the two dyes simultaneously. Three discrete neurons synapses at different observing planes within the RoI area are selected as the main structures. Compared with the direct wide-field fluorescent imaging results generated from the high-NA (NA=0.9) objective, the contours can be better distinctly mapped, which shows the desired extended DoF of the apochromatic SOL as proposed. Additionally, based on the flexible and controllable design method, the energy ratio for the focusing hotspot of the SOL can be further adjusted to satisfy various samples without damaging the biological activity, which greatly helps to build up the customized optical imaging system. Compared with the expensive confocal and multiphoton microscopes, the apochromatic SOL-based microscope can provide high-resolution and apochromatic property with extended DoF at low-cost and greater flexibility. Admittedly, the bare imaging results provided by the apochromatic SOL-based microscope system shows an inevitable background noise. To solve this problem, the noise reduction and template matching algorithms are employed to reconstruct the imaging results. It is believed that the results can be further improved by updating the used CCD camera with a higher quantum yield. In addition, the DoF for the apochromatic SOL can be extended to tens of micrometers by algorithm optimization.

The high-NA SOL with apochromatic extended DoF will promote the practical applications for the planar SOLs in many fields, such as non-invasive 3D biomedical imaging, laser tweezing, multicolor optical coherence tomography imaging, spectral microscopy imaging, lab-on-chip devices and micro/nanofabrication. The unique advantage of the customized light-field patterns for the SOLs in the far field will make biological imaging much more flexible and efficient.

**METHODS**

**GA-based optical field design.** In our theoretical design, genetic algorithm (GA), a powerful computational tool to solve multi-objective and multi-constraint optimization problems and generate Pareto-optimal solutions,[26, 47-49] is employed to optimize the SOL DoF while minimizing the side-lobe intensities and the main-lobe FWHM under monochromatic illumination. Since a subwavelength-sized optical field pattern can be constructed by employing an SOL consisting of multiple concentric microrings through the interference of their diffracted beams, in the GA-based SOL design, the outer radius, widths of the concentric rings, and locations of discrete foci along the optical axis are fixed for specific incident wavelengths while the central positions of the rings are set as variables. To produce a large and uniform DoF, the number of desired discrete axial focal spots $M$ are initially preset, and these focal hotspots are then gradually prolonged to overlap with each other; To reduce the side-lobe intensities and shrink the main-lobe FWHM, we restrain the field pattern's intensities along two orthogonal directions perpendicular to the optical axis. The intensity ratio between the side-lobes and the central focusing spot is set to be below 0.3 in order to produce a sharp contrast. This constraint ensures a high signal-to-noise ratio, favorable for fluorescent imaging by selectively illuminating biological tissues exclusively with the central focusing hotspot. With the aforementioned methodology, an optimization model is developed based on Eq. M1, which is subject to the conditions given in Eq. M2. The DoF, side-lobe intensity and main-lobe FWHM at different foci along the optical axis are simultaneously optimized by the three objective functions as follows:

$$G_1 = \max\left[I(0, f_{1-}; t_i; \lambda_q), I(f_{M+}, z; t_i; \lambda_q)\right];$$
$$G_2 = \max\left|I(f_{m-}, f_{m+}; t_i; \lambda_q) - I(f_{1-}, f_{1+}; t_i; \lambda_q)\right|;$$
$$G_3 = I\left(\frac{FWHM}{2}, f_{m-}; t_i; \lambda_q\right);$$
$$\min G_1, G_2, G_3$$

(Eq. M1)

*s.t.*

$$f_{n+} = f_{(n+1)-};$$
$$\begin{vmatrix} I(0, z_0; t_i; \lambda_q) & 1.15 \\ 1 & 1 \end{vmatrix} \leq 0 \leq \begin{vmatrix} I(0, z_0; t_i; \lambda_q) & 0.95 \\ 1 & 1 \end{vmatrix};$$
$$\begin{vmatrix} I(0, z_x; t_i; \lambda_q) & 0.3 \\ 1 & 1 \end{vmatrix} \leq 0;$$
$$\begin{vmatrix} I(r, f_m; t_i; \lambda_q) & 0.3 \\ 1 & 1 \end{vmatrix} \leq 0, \begin{vmatrix} r & k\frac{FWHM}{2} \\ 1 & 1 \end{vmatrix} \leq 0 \leq \begin{vmatrix} r & \frac{FWHM}{2} \\ 1 & 1 \end{vmatrix};$$
$$z_0 \in \left(f_1 - \frac{\Delta_f}{2}, f_M + \frac{\Delta_f}{2}\right);$$
$$z_x \in \left(0, f_1 - \frac{\Delta_f}{2}\right) \cup \left(f_n + \frac{\Delta_f}{2}, f_{n+1} - \frac{\Delta_f}{2}\right) \cup \left(f_M + \frac{\Delta_f}{2}, z\right);$$
$$q = 1, 2, \cdots, Q; m = 1, 2, \cdots, M; n = 1, 2, \cdots, M-1;$$
$$t_i \in \{0, 1\}, i = 1, 2, \cdots, N;$$

(Eq. M2)

where $I$ stands for the normalized field intensity of the needle-like focusing region; $\lambda_q$ is the incident wavelength; $f_{n+} = f_n + \Delta_f/2$, $f_{(n+1)-} = f_{n+1} - \Delta_f/2$; $f_{m-} = f_m - \Delta_f/2$ and $f_{m+} = f_m + \Delta_f/2$, with $f_m$ being the focal distance of the $m$-th hotspot and $\Delta_f$, the DoF of each hotspot; $Q$ is the total number of incident wavelengths of interest; $t_i$ is the transmittance value of the $i$-th annular ring and $N$ represents the total number of concentric rings in the mask. The first objective function $min(G_1)$ seeks the energy surrounding each main axial focal spot to be as low as possible, with the aim to ensure the maximum intensity at the main focal spot; the second objective function $min(G_2)$ minimizes the difference in the intensities of all hotspots; the third objective function $min(G_3)$ minimize the FWHM of the main lobe, ensuring high resolution. Additionally, in order to generate a highly uniform main lobe, the normalized intensity over the focusing region is set to be within 0.95 and 1.05, and the axial intensity of other secondary points is set to be below 0.3 to ensure a sharp intensity contrast between the side and main lobes. To construct an unconstrained optimization process, the intensity modulation of the needle-like focusing pattern formed by merging adjacent axial multifoci is added as a penalty function term. To accelerate the numerical calculation, a fast Hankel transform algorithm is employed to ensure high accuracy, fast optimization, and less data storage.[50] Five main axial hotspots ($M = 5$) are chosen in this work, giving rise to the phase distribution profile of the optimized SOL shown in Fig. S2.

Note that the optimization at three separate wavelengths, as well as the extension of DoF (10λ) achieved in this work does not represent the ultimate potential of our GA-based optimization method. An even longer DoF of 23.4λ has been realized at 640 nm as unveiled in Fig. S3, with its optical characteristics shown in Tab. S3 in the Supplementary Information.

**Wafer-scale nanofabrication.** Parallel fabrication of optical elements, the main trend fabrication method compared to the complex and time-consuming processes such as electron beam lithography (EBL) and focused ion beam milling (FIB),[18, 28, 30-32, 35] has been employed in our work owing to its micrometric feature size. For the sake of practical application of the planar SOLs integrated with the on-chip optical imaging system, a functionalized SOL with 1-mm diameter is efficiently fabricated in a mass production way. The 2.5-μm feature size of the annuli of our optimized SOL makes it possible to fabricate the millimeter-scale device through the conventional optical lithography process.[36, 37] The SixNy film is deposited by PECVD, followed by optical lithography used for graphic processing and RIE for etching. The fabrication process and the optical refractive index of SixNy film are presented in Supplementary Information Note 2 and Fig. S4.

To design a realistic dielectric apochromatic lens, the material dispersion induced phase modulation difference (>π for short wavelengths, <π for longer wavelengths) should be taken into consideration in the optimization process. For example, with an etching depth of 213 nm, a step in the $Si_xN_y$ layer creates a phase retardation of 1.1π at $\lambda_B$=488 nm and 0.8π at $\lambda_R$=640 nm. The material dispersion related phase deceleration is included as additional phase in our GA optimization process.

**Splicing of fluorescent-labeled cell images.** Different from the images captured of the resolution test target in high contrast, the imaging process of the cell fluorescent images is challenging especially for the feature matching. In this part, we make an introduction of the feature matching and image fusion process of the sequence images. Since affine transformation will occur between sequence images, we refer to the method of panoramic image Mosaic[51] to estimate affine matrix. Compared with the extraction effects of ORB,[52] FAST,[53] SURF[54] and other key points, SIFT key points[55] show excellent translation and rotation invariant properties. Henceforth, SIFT key points algorithm are chosen to complete the image matching so that the cell texture features could be clearly reconstructed. The KD-Tree[56] is applied to select the most appropriate matching objects, in $O(n \log n)$ time, where the SIFT points and the affine matrix *H* is estimated.[51]

$$H_{ij} = K_i R_i R_j^T K_j^{-1} \quad \text{(Eq. M3)}$$

where the camera parameter matrices $K$ and $R$ are defined as follows

$$K_i = \begin{bmatrix} f_i & 0 & 0 \\ 0 & f_i & 0 \\ 0 & 0 & 1 \end{bmatrix} \quad \text{(Eq. M4)}$$

$$R_i = e^{[\theta_i]_X}, [\theta_i]_X = \begin{bmatrix} 0 & -\theta_{i3} & -\theta_{i2} \\ -\theta_{i3} & 0 & -\theta_{i1} \\ -\theta_{i2} & \theta_{i1} & 0 \end{bmatrix} \quad \text{(Eq. M5)}$$

To improve the accuracy of the affine transformation matrix $H$ estimation, RANSAC[57] algorithm is used to estimate the interior points in the matched SIFT point pairs. Then Levenberg-Marquardt algorithm is utilized to calculate the least square fitting for multiple point pairs (greater than 4) to get the transformation matrix $H$. Referring to the methods mentioned in[51] and,[58] the spatial consistency of multiple images is explored and the accumulated errors are reduced. In particular, the Bundle Adjustment method is adopted to process all images simultaneously, and the updated equation (Eq. M4) is iteratively solved to obtain a more accurate transformation matrix.

$$\Phi = (J^T J + \lambda C_p^{-1})^{-1} J^T r \quad \text{(Eq. M6)}$$

where $\Phi$ are all the parameters, $r$ the residuals and $J=\partial r/\partial \Phi$, the (diagonal) covariance matrix $C_p$ is

$$C_p = \begin{bmatrix} \sigma_\theta^2 & 0 & 0 & 0 & 0 & \cdots \\ 0 & \sigma_\theta^2 & 0 & 0 & 0 & \cdots \\ 0 & 0 & \sigma_\theta^2 & 0 & 0 & \cdots \\ 0 & 0 & 0 & \sigma_f^2 & 0 & \cdots \\ 0 & 0 & 0 & 0 & \sigma_0^2 & \cdots \\ \vdots & \vdots & \vdots & \vdots & \ddots & \cdots \end{bmatrix} \quad \text{(Eq. M7)}$$

Considering the environmental vibration and laser instability which may cause the difference in illumination conditions for each picture, a sharp gradient changes is noticeable in some areas during the fusion. Gain compensations are therefore introduced, resulting in the error function (Eq. M6) to be constructed as the sum of the gain times and the normalized intensity errors of all overlapped pixels.

$$e = \frac{1}{2}\sum_{i=1}^{n}\sum_{j=1}^{n}\sum_{\substack{u_i \in \mathcal{R}(i,j) \\ \tilde{u}_i = H_{ij}\tilde{u}_j}} (g_i I_i(u_i) - g_j I_j(u_j))^2 \quad \text{(Eq. M8)}$$

where $g_i$, $g_j$ are the gains, and $\mathcal{R}(i,j)$ is the overlapped region between images $i$ and $j$. Here a variable coefficient average method is applied to replace the same mean for the global images. To obtain the final fusion results, it is the necessary to track the central spot and enhance the clarity of the scanned texture of the central spot. The sub-regional average approximation is adopted here, $I_i(u_i)$ can be approximated by $\bar{I}_{ij}$:

$$\bar{I}_{ij} = \frac{\sum_{u_i \in \mathcal{R}(i,j)} I_i(u_i)}{\sum_{u_i \in \mathcal{R}(i,j)} \alpha}, \alpha = \begin{cases} 0.8, I(\mathcal{R}(i,j)) > 0.85\max\{I(\mathcal{R}(i,j))\} \\ 1, I(\mathcal{R}(i,j)) \leq 0.85\max\{I(\mathcal{R}(i,j))\} \end{cases} \quad \text{(Eq. M9)}$$

**Data availability**

The authors declare that all the data supporting the findings of this study are available within this paper and its Supplementary Information file, or available from the corresponding author upon reasonable request.

**Acknowledgement**

We acknowledge the financial support by the Science, Technology and Innovation Commission of Shenzhen Municipality (JCYJ20180508151936092), the National Natural Science Foundation of China (51975483), the Key Research Projects of Shaanxi Province (2020ZDLGY01-03), the Ningbo Natural Science Foundation (202003N4033), the Open Research Fund of CAS Key Laboratory of Spectral Imaging Technology (LSIT201912W), the Innovation Foundation for Doctor Dissertation of Northwestern Polytechnical University (CX201908), and the City University of Hong Kong (9610456).

**Notes**

All other authors declare they have no competing interests.

**Additional Information**

Supplementary material for this article is available.